\documentclass[twocolumns]{aa}
\usepackage{graphicx}
\usepackage{epsfig,graphicx,rotating,portland,longtable,natbib}
\usepackage{txfonts}
\def\gs{{_>\atop^{\sim}}}

\begin{document}
\title{Probing the complex environments of GRB host galaxies and intervening systems: high resolution 
spectroscopy of GRB050922C%
\footnote{Based on observations collected at the European Southern
Observatory (ESO) with the VLT/Kueyen telescope, Paranal, Chile, in the
framework of program 075.A-0603.}}

%   \subtitle{}

\author{S. Piranomonte\inst{1}
P. A. Ward\inst{2}
F. Fiore\inst{1}
S. D. Vergani\inst{3,4,5}
V. D'Elia\inst{1}
Y. Krongold\inst{6}
F. Nicastro\inst{1}
E. J. A. Meurs\inst{4}
G. Chincarini \inst{5,7}
S. Covino \inst{5}
M. Della Valle \inst{8,9,10}
D. Fugazza \inst{5}
L. Norci \inst{3}
L. Sbordone \inst{11}
L. Stella \inst{1}	
G. Tagliaferri \inst{5}
D. N. Burrows \inst{12}
N. Gehrels \inst{13}
P. Goldoni \inst{14,15}
D. Malesani \inst{16}
I. F. Mirabel \inst{17}
L. J. Pellizza \inst{18}
R. Perna \inst{19}
}

\institute{INAF-Osservatorio Astronomico di Roma, 
via Frascati 33, 00040 Monte Porzio Catone (RM),  Italy.
\email{piranomonte@oa-roma.inaf.it}
\and 
%2
Mullard Space Science Laboratory, University College London, Holmbury St. Mary, Dorking Surrey, RH56NT, UK.
\and
%3
Dublin Institute for Advanced Studies, 31 Fitzwilliam Place - Dublin 2, Ireland.
\and
%4
School of Physical Sciences and NCPST, Dublin City University, Glasnevin, Dublin 9, Ireland.
\and
%5
INAF-Osservatorio Astronomico di Brera, via E. Bianchi 46, 23807 Merate (LC), Italy.
\and
%6
Instituto de Astronomia, Universidad nacional Autonoma de Mexico, Apartado Postal 70-264, 04510 Mexico DF, Mexico.
\and
%7
Universit\`a degli Studi di Milano Bicocca, Piazza della Scienza 3, 20126 Milano, Italy.
\and
%8
INAF-Osservatorio Astronomico di Capodimonte, Salita Moiariello 16, 80131 Napoli, Italy.
\and
%9
International Center for Relativistic Astrophysics Network, 65122, Pescara, Italy.
\and
European Southern Observatory, Karl-Schwarzschild-Strasse 2, D-85748 Garching bei M\"unchen, Germany.
\and
%10
GEPI, Observatoire de Paris, CNRS, Universit\'e Paris Diderot, Place Jules Janssen, 92190 Meudon, France.
\and
%11
Department of Astronomy and Astrophysics, Pennsylvania State University, University Park, Pennsylvania 16802, USA.
\and
%12
NASA-Goddard Space Flight Center, Greenbelt, Maryland, 20771, USA.
\and
%13
APC, Laboratoire Astroparticule et Cosmologie, UMR 7164, 11 Place
Marcelin Berthelot, 75231 Paris Cedex 05, France.
\and
%14
CEA Saclay, DSM/DAPNIA/Service d'Astrophysique, 91191 Gif-sur-Yvette,
France.
\and
%15
Dark Cosmology Centre, Niels Bohr Institute, Juliane Maries Vej 30, 2100 K\o{}benhavn \O, Denmark.
%\and
%Mullard Space Science Laboratory, University College London, Holmbury St. Mary, Dorking Surrey, RH56NT, UK 
\and
%16
European Southern Observatory, Alonso de C\'ordova 3107, Santiago 19, Chile (on leave from CEA Saclay, France).
%\and
 %INAF-Istituto di Astrofisica Spaziale e Fisica Cosmica, Via Bassini, 15, I-20133, Milano, Italy.
 \and
 %17
Instituto de Astronom\'{\i}a y F\'{\i}sica del Espacio, CONICET/UBA, Casilla de Correos 67, Suc. 28, (1428) BuenosAires, Argentina. 
\and
%18
JILA, Campus Box 440, University of Colorado, Boulder, CO 80309-0440, USA.
}

\date{August 25, 2008}

\abstract % context heading (optional)  leave it empty if necessary 
{} 
% aims heading (mandatory) 
{The aim of this paper is to investigate the environment of gamma ray
bursts (GRBs) and the interstellar matter of their host galaxies.}
% methods heading (mandatory) 
{We use to this purpose high resolution spectroscopic observations of the afterglow of 
GRB050922C, obtained with UVES/VLT $\sim3.5$ hours after the GRB
event.}
% results heading (mandatory)
{We found that, as for most high resolution spectra of GRBs, the
spectrum of the afterglow of GRB050922C is complex.  At least seven components
contribute to the main absorption system at $z=2.1992$. The detection
of lines of neutral elements like MgI and the detection of
fine-structure levels of the ions FeII, SiII and CII allows us to
separate components in the GRB ISM along the line of sight. 
Moreover, in addition to the main system, we have analyzed the five intervening systems between $z = 2.077$ and $z =
1.5664$ identified along the GRB line of sight.}
% conclusions heading (optional), leave it empty if necessary
{GRB afterglow spectra are very complex, but full of information.
This can be used to disentangle the contribution of the different
parts of the GRB host galaxy and to study their properties. Our
metallicity estimates agree with the scenario of GRBs exploding in low
metallicity galaxies}
\keywords{gamma rays: bursts - cosmology: observations - galaxies:
abundances - ISM} 

\authorrunning {Piranomonte et al.}  
\titlerunning{UVES/VLT high resolution spectroscopy of GRB050922C}

\maketitle

%________________________________________________________________

\section{Introduction}

Soon after the discovery of the cosmological origin of gamma-ray
bursts (GRBs) it was realized that they provide powerful tools to
investigate the high redshift Universe. Minutes to hours after a GRB,
its optical afterglow can sometimes be as bright as magnitude 13--18,
such that it can be used as a bright beacon to gather high resolution
(a few tens of km\,s$^{-1}$ in the optical band), high quality (signal
to noise $>10$ per resolution element) spectra. Absorption lines
produced by the matter along the line of sight are superimposed on
these spectra and allow a detailed investigation of the medium
surrounding the GRB and of the host galaxy interstellar matter (ISM).

As of today, about a dozen GRBs have been observed with high
resolution spectrographs (and less than one hundred with standard low
resolution spectrographs, see \citealt{Savaglio06} and references
therein).  These high resolution spectroscopic observations showed
that the absorption systems along GRB sightlines are complex, with
many components. They are resolved down to a width of a few tens
km\,s$^{-1}$ and contribute to each main absorption system, spanning a
total velocity range of typically hundreds km s$^{-1}$. In one
case, GRB021004, all the systems likely to be associated to the GRB host
galaxy span a velocity range of a few thousands km s$^{-1}$, although \citet{Chen07} suggest that
 the high velocity 3000 km s$^{-1}$ CIV component may no not be related to the GRB host galaxy. Both low
and high-ionization lines are normally observed, as well as fine
structure lines of CII, SiII, OI, FeII etc. This is at odds with what
is observed from high resolution spectroscopy of QSO ``associated''
absorption lines (AALs; \citealt{dodorico04}). First, the equivalent
widths and the inferred optical depths along GRB sightlines are much
greater than both QSO AALs and those associated with galaxies along
QSO sightlines. Second, despite more than 30 years of investigations,
only sparse detections of fine-structure lines are available along
QSOs sightlines due to a lower density environment around them. 
Fine structure lines have been detected only in Broad Absorption Line
QSO spectra \citep{Srianand01, Srianand00, Hall02}.

The main reasons for these marked observational differences are
probably linked to the characteristics of the phenomena and to their
host galaxies. QSOs are long-lasting sources that ionize, partially or
totally, the gas over their entire host galaxy. Instead, GRBs are
transient phenomena, affecting only smaller regions, close to the site
of the GRB explosion. Moreover, the star-forming regions surrounding
GRBs are probably much richer in enriched gas than the giant
elliptical galaxies, which are the typical host of luminous QSOs
\citep{Savaglio06}.  On the other hand, faint metal line systems along
QSO sightlines probe mainly galaxy haloes, rather than their bulges or
discs.  Therefore, GRBs constitute a complementary tool to investigate
the ISM of typical high redshift galaxies.

After the first pioneering programs that used BeppoSAX and HETE2
triggers to obtain high resolution spectroscopy of the optical
afterglows 12-24 hours after the bursts \citep[e.g.][]{fiore05}, we
are now in a ``golden age'' for this kind of studies. This is because
of the GRB-dedicated Swift satellite which provides accurate GRB
positions on time scales as short as tens of seconds, and the rapid
response mode (RRM) of the ESO/VLT. Using this new revolutionary
observing mode it is possible to obtain GRB afterglow spectra from an
8m class telescope with delays from $\sim$10\,min down to a few hours
after the GRB event.

The absorption systems along GRB sightlines can be divided into three
broad categories: (1) systems associated with the GRB surrounding
medium. The physical, dynamical and chemical state of the medium in
the star-forming region hosting the GRB progenitor can be modified by
the GRB (long type), through shock waves and ionizing photons. Strong
fine structure lines are always observed in these systems.  Physical
and geometrical parameters for the absorbers can be derived comparing
the observed line ratios to those predicted by ionization codes
\citep[e.g.][]{Nicastro99, Perna02, Lazzati06, vrees07}. (2) Systems
associated with random clouds of the ISM of the host galaxy along the
line of sight. These systems show low ionization lines and ground
state lines, implying distances from the GRB explosion site larger
than a few hundred pc.  (3) Systems originating in the intergalactic
matter along the line of sight.

The distinction between the first and second category is not always
clear cut.   \citet{Dessauges06} analyzed the afterglow spectra of GRB 020813 obtained in two epochs. They report a decline by a factor of 5 in the equivalent width of the Fe II $\lambda$2396 transition, giving an estimate on its distance from the afterglow of about 50-100 pc.
\citet{vrees07} discovered large variations of FeII fine-structure lines in the spectra of GRB060418 on rest frame time
scales of a fraction of an hour, and interpreted them as due to
ultraviolet (UV) pumping from the afterglow radiation field. Their
detailed modeling of the observed variability suggests a distance of
the FeII absorbing cloud from the GRB explosion site of $\sim1.7$~kpc,
comparable to the size of a typical galaxy at $z \sim 1.4$.  This
would imply that, at least in this case, the entire galaxy (or a
significant fraction of it) is affected by the GRB explosion. 
Recently \citet{delia08} analyzed the high resolution spectra of GRB080319B afterglow 9 minutes and 8 hours after the GRB onset. They found the strongest Fe II fine structure lines ever observed in a GRB. A few hours later the optical depth of these lines was reduced by a factor of 4-20, and the optical/UV flux by a factor of $\sim$60. This proves that the excitation of the observed fine structure lines is due to ``pumping'' by the GRB UV photons. A comparison of the observed ratio between the number of photons absorbed by the excited state and those in the Fe II ground state suggests that the six absorbers are $\gs18-34$ kpc from the GRB site.
%Therefore \citet{prochaska08} demonstrated that GRB afterglows photoionize nitrogen to NV at r$\sim10$ pc. In this scenario, the observations imply the progenitor's stellar wind is confined to r$<$10 pc which suggests the GRB progenitors occur within dense (n $> 10^3$ cm$^{-3}$) environments, typical of molecular clouds.}

Our goal is to better disentangle different absorption systems by
using Ultra-violet and Visual Echelle Spectrograph (UVES;
\citealt{dekker00}) observations of GRB050922C. The burst was detected
by Swift on 2005 September 22 19:55:50 UT \citep{Norris05} and, after
4~s, by Konus/Wind \citep{Golenetskii05}. ROTSE discovered a bright
transient source at RA(J2000$)=21^{\rm h}09^{\rm m}33\fs1$ and
DEC(J2000$)=-08\degr45\arcmin29\farcs8$, 172.4~s after the burst,
which was promptly identified as the optical afterglow of GR050922C
\citep{Rykoff05} with a magnitude $R = 14.7$. The redshift of the GRB,
$z=2.1992$, was determined by \citet{Jakobsson05} using ALFOSC at the
Nordic Optical Telescope at the Observatorio del Roque de los
Muchachos in La Palma. This redshift was confirmed shortly after
through TNG/DOLoRes spectroscopy \citep{piranomonte05}. The afterglow
was then observed by UVES/VLT at high spectroscopic resolution ($R
\sim 40000$) about 3.5 hours after the burst, when its magnitude was
$R=19.0$.  The spectrum of GRB050922C is especially rich of features
and contains clear examples of the three system categories discussed
above. In addition to the main system at z=2.1992, there are five 
intervening systems at $z=1.5664$, 1.9891, 2.0087, 2.077, 2.1421
(confirmed also by \citealt{Sudilovsky07}).
 
The system associated with the host galaxy presents several components, some
showing strong high-ionization lines and strong fine-structure lines,
 some showing only low-ionization transitions, thus suggesting that
we detect in this GRB both the circum-burst medium and random clouds
of the host galaxy ISM along the line of sight.

This paper is organized as follows. Section 2 presents the
observations and data reduction; Section 3 gives a brief summary of
all absorbing systems identified in the UVES spectrum; Section 4
presents a detailed analysis of the absorption features belonging to
the main absorption system at $z=2.1992$; finally, in Section 5 the
results are discussed and conclusions are presented.

\begin{figure*}
\begin{center}
\includegraphics[height= 8cm, width=6cm, angle=-90]{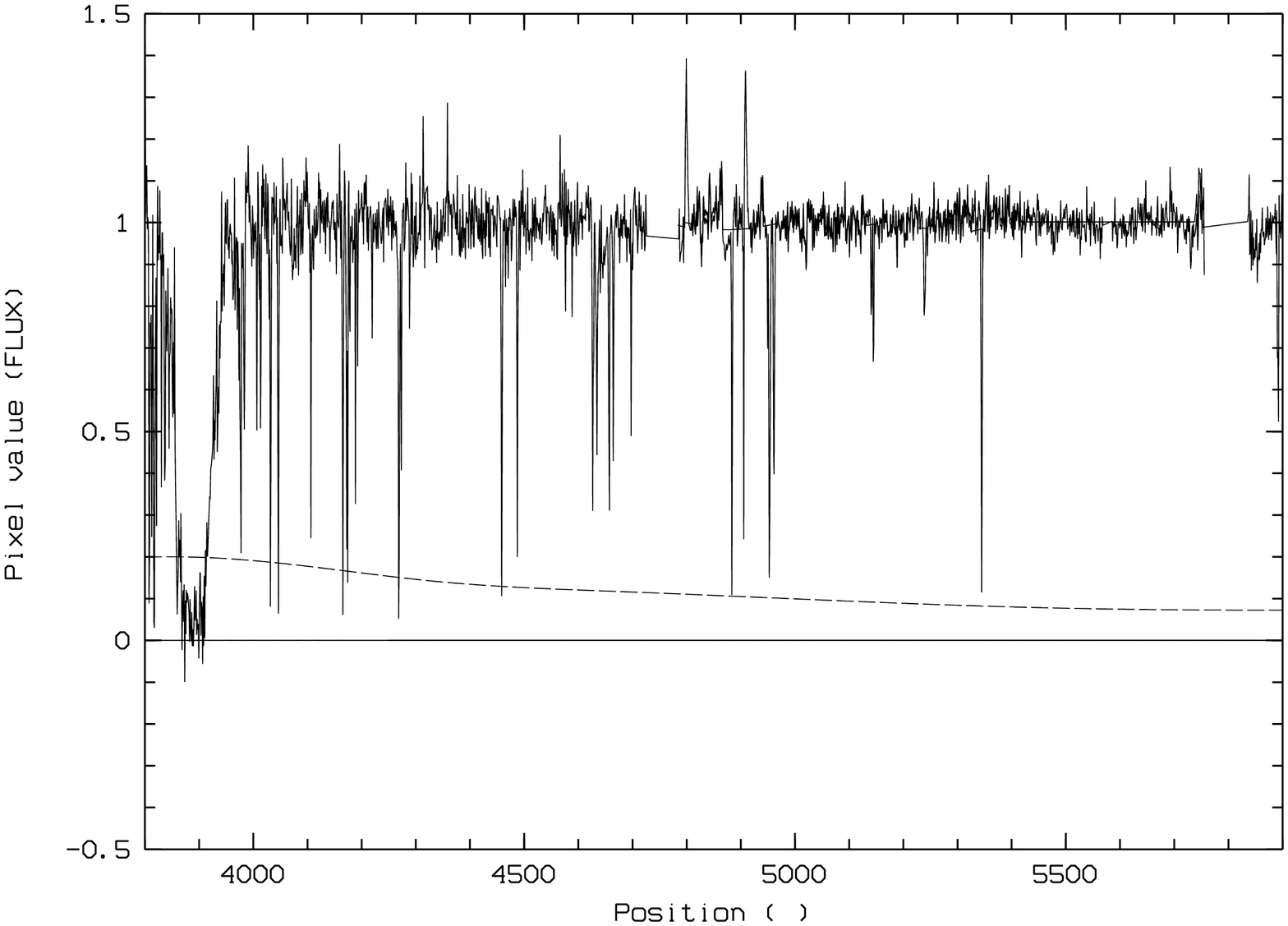}
\includegraphics[height=8cm, width=6cm, angle=-90]{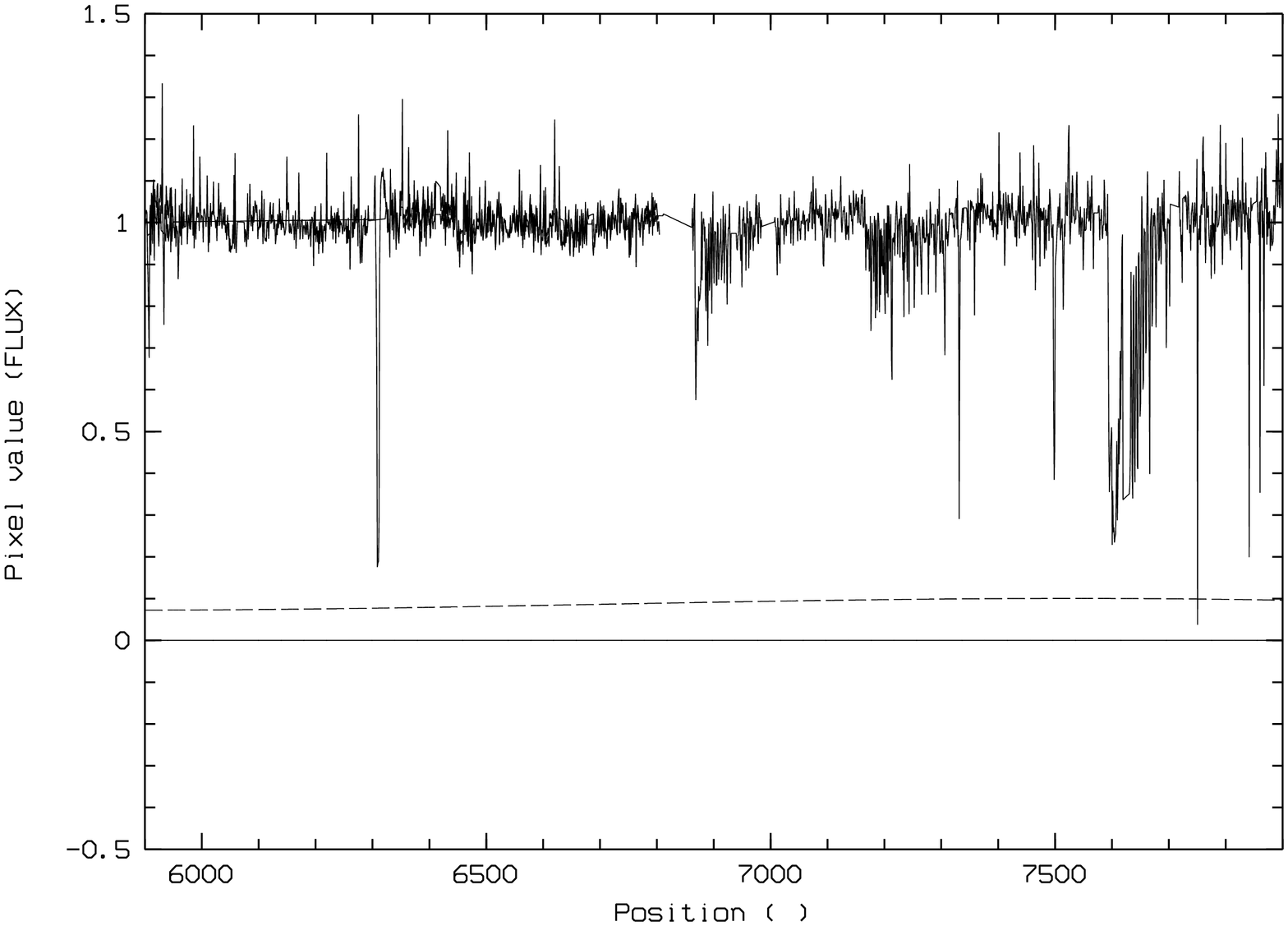}
\includegraphics[height=8cm, width=6cm, angle=-90]{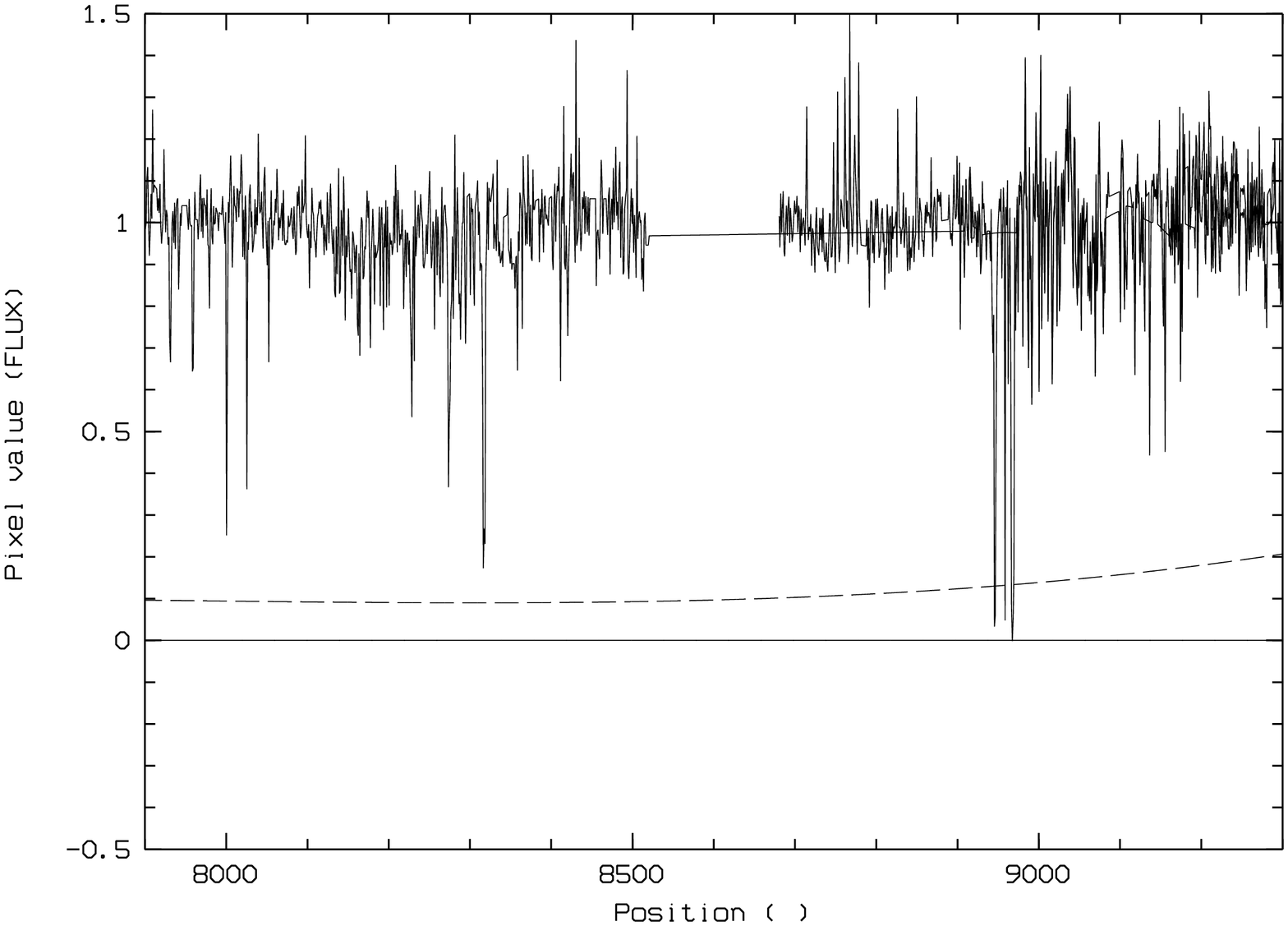}
\end{center}
\caption{The full high resolution spectrum of GRB050922C. Dashed lines represent
the noise spectrum.}
\label{ove}
\end{figure*}

\section{Observations and data analysis}

\subsection{VLT observations}

In the framework of ESO program 075.A-0603 we observed the afterglow
of GRB050922C with UVES mounted at the VLT-UT2 telescope.  Table
\ref{obs_log} gives the log of the observations.  Both UVES dichroics,
as well as the red and the blue arms, were used.  This allowed us to
achieve a particularly wide spectral coverage, extending from
$\sim3200$~\AA{} to $\sim 10000$~\AA. In order to maximize the signal
to noise ratio the CCD was rebinned in $2\times2$ pixels. The data
reduction was carried out by using the UVES pipeline
\citep{ballester00}. The final useful spectra extend from about
3750~\AA{} to about 9780~\AA{}. Following \citet{fiore05} the
total spectrum was rebinned to 0.1~\AA~  to increase the
signal-to-noise ratio. This resolution is still good enough to sample
with at least 3-4 bins weak lines. The resolution element, set to two
pixels, ranges then from 14 km\,s$^{-1}$ at 4200~\AA{} to 6.6
km\,s$^{-1}$ at 9000~\AA.  The noise spectrum, used to determine the
errors on the best fit line parameters, was calculated from the 
real-background-subtracted and rebinned spectrum using line-free
regions. This takes into account both statistical errors and
systematic errors in the pipeline processing and background
subtraction.

%the high resolution UV-visual echelle spectrograph (UVES, Dekker et al. 2000)

\subsection{Column densities}
\label{fit}

The line fitting was performed using the MIDAS package FITLYMAN
\citep{Fontana95}.  This uses a Voigt profile and yields independently
the column density $N$ and the Doppler parameter $b$ for each absorption
component.  For each absorption system several lines, spread over the
entire spectral range covered by the UVES observations, were fitted
simultaneously, using the same number of components for each line, and
the same redshift and $b$ value for each component.

In all of the following figures showing plots from FITLYMAN
(Fig. \ref{2_14} - \ref{FeII}) the x-axis represents the velocity
shift of the gas with respect to the zero, the y-axis represents the flux
normalized to 1 for each element, the horizontal blue lines and dashed
blue lines represent the flux level of the spectrum normalised to 1
and zero point respectively. The red line corresponds to the fit
profile of the individual components and the magenta and green lines
correspond to the polynomial approximation of the noise and residual
spectrum respectively.

\begin{table*}
\caption{Journal of observations.}
\label{obs_log}
\tiny{
\begin{tabular}{lccccccc}
%\tableline\tableline
%da completare
\hline
Date & Dichroic &\multicolumn{2}{c}{Central wavelength} & Slit width & Seeing    & Exposure & Time since GRB \\
(UT) &          & Blue arm   & Red arm                  & (\arcsec)  & (\arcsec) & (min)    & (hr)           \\
\hline
09/22/05 00:07:17 & 1	& 3460 \AA & 5800 \AA & 1 & $\la1$ & 50 & 4.15 \\
09/22/05 01:02:09 & 2	& 4370 \AA & 8600 \AA & 1 & $\la1$ & 50 & 5.07 \\
%\tableline
\hline
\end{tabular}
}
\end{table*}

\section{Summary of the absorption systems}

The high resolution spectrum of GRB050922C (see Fig. \ref{ove})
exhibits a large number of absorption features, belonging to different
absorption systems. The main system is due to the gas inferred to be
in the host galaxy, at redshift $z=2.1992$. We identified five more
intervening absorbers at $z < 2.1992$ ($z=1.5664$, 1.5691, 1.9891,
2.0087, 2.077 and 2.1416).  \citet{Sudilovsky07} found one more
intervening systems $z=1.6911$, but we could not confirm that this is
real because the {\mbox{\ion{C}{IV}}} doublet is coincident with the
SiII$\lambda$1304 and OI$\lambda$1302, OI*$\lambda$1304 of the main
system at $z=2.1992$.

Each system is briefly illustrated in the next subsections.

\subsection{The host galaxy at $z=2.1992$}

This is the system with the highest number of features in the UVES
spectrum. Fig. \ref{ove} shows the Ly-$\alpha$ absorption at $\sim
3870$~\AA. The redshift inferred from this feature is confirmed by a
high number of metal-line transitions, the narrowest of which allow a
more precise determination: $z=2.1992 \pm 0.0005$.  This system exhibits a
large number of absorption lines, from the neutral hydrogen Ly-$\alpha$ 
and neutral metal-absorption lines (OI, NI
and FeI) to low-ionization lines ({\mbox{\ion{C}{II}}},
{\mbox{\ion{Si}{II}}}, {\mbox{\ion{Mg}{II}}}, {\mbox{\ion{Al}{II}}},
{\mbox{\ion{Fe}{II}}}, {\mbox{\ion{Al}{III}}}) and high-ionization
absorption features ({\mbox{\ion{C}{IV}}}, {\mbox{\ion{Si}{IV}}},
{\mbox{\ion{N}{V}}}, {\mbox{\ion{P}{V}}}, {\mbox{\ion{O}{VI}}}). In
addition, strong fine-structure lines ({\mbox{\ion{C}{II*}}},
{\mbox{\ion{Si}{II*}}}, {\mbox{\ion{O}{I*}}}, {\mbox{\ion{O}{II**}}},
{\mbox{\ion{Fe}{II*}}}, {\mbox{\ion{OI}{II**}}}) have been identified.
The total velocity range spanned by this system is $\sim210$ km\,s$^{-1}$.
This system is discussed in detail in the next sections.

%%%%%%%%%%%%%%%%%%%%%%%%%

%%%%%%%new part of intervening systems %%%%%%%%%%%%%%

\subsection{The intervening absorber at $z = 2.1416$}

\begin{figure}[h]
\centering
%\begin{center}
\includegraphics[height=7cm, width=7cm]{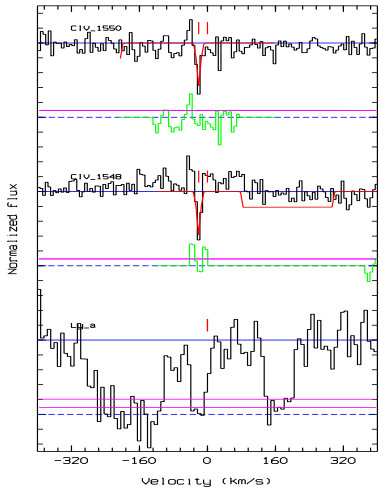}
%\end{center}
\caption{The intervening absorber at $z = 2.1416$. The zero of the
velocity scale refers to this redshift.}
\label{2_14}
\end{figure}

The intervening system at $z=2.1416$ shows one main component spanning a
velocity range of $\sim$ 20 km\,s$^{-1}$ (see Fig. \ref{2_14}). This system
presents HI and the CIV transition.  
The total HI column density is logN(HI)=18.1$\pm$0.1 cm$^{-2}$.

The derived column density of CIV is given in Table \ref{table_2.14}.

\begin{table}[h]
\def~{\hphantom{0}}

{\tiny

\caption{Column densities for the $z=2.1416$ intervening system.}
\begin{center}
\begin{tabular}{cc}\hline

\bf Transition   &  2.1416  \\\hline
%SiII & 12.54$\pm$2.91\\
%\bf HI  & 18.1 $\pm$0.1 \\
%CII  & 13.91$\pm$0.07 \\
CIV & 13.29$\pm$0.07 \\\hline
\end{tabular}

\label{table_2.14}
\end{center}}
\end{table}

\subsection{The intervening absorber at $z = 2.077$}

\begin{figure}[h]
\centering
%\begin{center}
%\includegraphics[height=8cm, width=9cm]{int_1_2.077.ps}
\includegraphics[height=8cm, width=7.6cm]{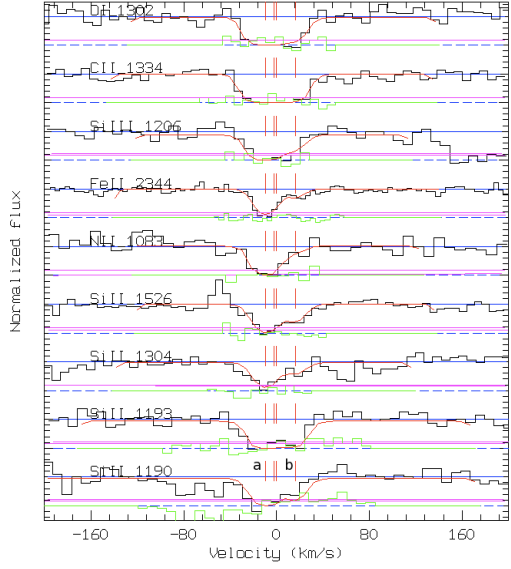}
\includegraphics[height=6cm, width=9cm]{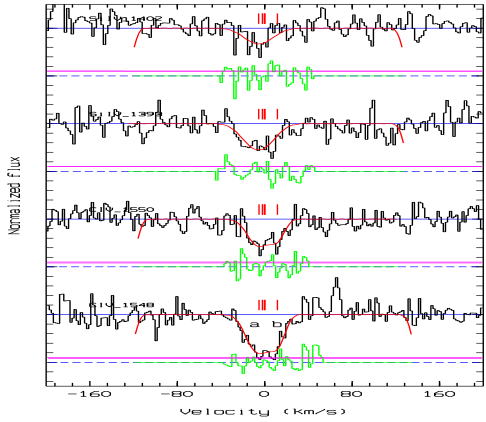}
%\end{center}
\caption{The intervening absorber at $z = 2.077$. The zero of the
velocity scale refers to this redshift.}
\label{2_1}
\end{figure}

The intervening system at $z=2.077$ shows two main components
spanning a velocity range of $\sim 50$ km\,s$^{-1}$ (see
Fig. \ref{2_1}). This system presents HI and neutral metal-absorption lines
(OI, NI), many low-ionization lines (FeII, SiII, NII, CII)
and some high-ionization transitions (SiIII, AlIII, SiIV, CIV). The derived
column densities are given in Table \ref{table_2.07}. We could not separate the two components for HI. 
 The total HI column density is logN(HI)=20.3$\pm$0.3 cm$^{-2}$.

\begin{table}[h]
\def~{\hphantom{0}}

{\tiny
\begin{center}
\caption{Column densities for the $z=2.077$ intervening system.}
\begin{tabular}{ccc}
\hline
\bf Transition  & 2.077a        & 2.077b \\
\hline

FeII & 14.45$\pm$0.08& 13.06$\pm$0.04 \\
SiII & 14.18$\pm$0.08& 13.54$\pm$0.07 \\
NII  & 14.56$\pm$0.06& 13.26$\pm$0.06 \\
%MgII &       -       & 19.0$\pm$0.5 \\
CII  & 15.7$\pm$0.3& 14.3$\pm$0.3 \\
SiIII& 13.9$\pm$0.6& 12.5$\pm$0.3 \\
OI   & 15.6$\pm$0.3& 14.4$\pm$0.2 \\
NI   & 14.2$\pm$0.1& 13.4$\pm$0.2 \\
%TiII & 13.9$\pm$0.1& 13.5$\pm$0.1 \\
AlIII&       -       & 12.2$\pm$0.1 \\
SiIV & 13.05$\pm$0.09& 12.5$\pm$0.2 \\
CIV & 13.70 $\pm$0.06& 13.3$\pm$0.1 \\
\hline
\end{tabular}

\label{table_2.07}
\end{center}}
\end{table}

\subsection{The intervening absorber at $z = 2.0081$}

\begin{figure}[h]
%\begin{center}
\centering
\includegraphics[height=7cm, width=8cm, angle=-90]{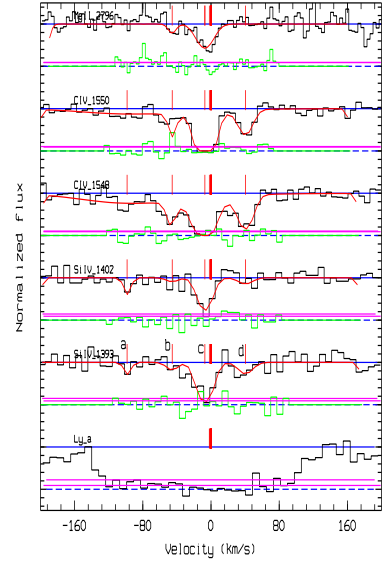}
%\end{center}
\caption{The intervening absorber at $z = 2.0081$. The zero of the
velocity scale refers to this redshift.}
\label{2_0}
\end{figure}

The intervening system at $z=2.0081$ shows four main components spanning
a velocity range of $\sim 140$ km\,s$^{-1}$ (see Fig.\ref{2_0}). This system
presents  HI, one low-ionization line (MgII) and some high-ionization
transitions (SiIV, CIV). The derived column densities are given in
Table \ref{table_2.0}.
We could not separate the four components for HI. The 
total HI column density is log $N$(HI)=18.2$\pm$0.2 cm$^{-2}$.

\begin{table}[h]
\def~{\hphantom{0}}
{\tiny
\begin{center}
\caption{Column densities for the $z=2.0081$ intervening system.}
\begin{tabular}{ccccc}
\hline
\bf Transition   &  2.0081a & 2.0081b & 2.0081c & 2.0081d \\
\hline
SiIV &12.4$\pm$0.2& 12.2$\pm$0.3 & 13.5$\pm$0.1 & 12.5$\pm$0.2\\
CIV  &13.1$\pm$0.1& 13.5$\pm$0.1 & 14.8$\pm$0.1 & 13.7$\pm$0.1 \\
MgII & - & 12.0$\pm$0.2 & 12.64$\pm$ 0.04 & - \\
\hline
\end{tabular}
\label{table_2.0}
\end{center}}

\end{table}

\subsection{The intervening absorber at $z = 1.9885$.}

\begin{figure}[h]
%\begin{center}
\centering
\includegraphics[height=7cm, width=9cm]{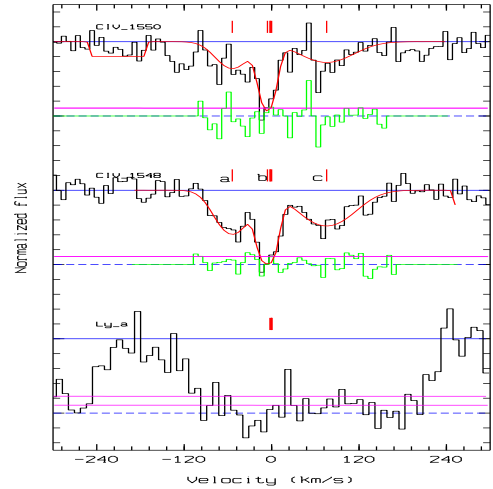}
%\end{center}
\caption{The intervening absorber at $z = 1.9885$. The zero of the
velocity scale refers to this redshift.}
\label{1_9}
\end{figure}

The intervening system at $z=1.9885$ shows three main components:
$z=1.9885$a, 1.9885b and 1.9885c, spanning a velocity range of $\sim
120$ km\,s$^{-1}$ (see Fig. \ref{1_9}). It presents HI and
high-ionization lines (SiIV, CIV). We could not constrain the
SiIV 1393 transition because it is blended with the saturated OI 1302
line from the GRB host galaxy. The derived column densities are given
in Table \ref{tab_1.9}.  We could not separate the three
components for HI. The total HI column density is log$N$(HI)=18.2
$\pm$0.4 cm$^{-2}$.

\begin{table}[h]
\def~{\hphantom{0}}
{\tiny
\begin{center}
\caption{Column densities for the $z=1.9885$ intervening system.}
\begin{tabular}{ccccc}
\hline
\bf Transition   &  1.9885a & 1.9885b & 1.9885c \\
\hline
CIV  &13.78$\pm$0.05&14.2$\pm$0.1& 13.79$\pm$0.04 \\
\hline
\end{tabular}

\label{tab_1.9}
\end{center}}
\end{table}

%%%%%%%%%%%%%%%%%%%%%%

\subsection{The intervening absorbers at $z = 1.5691$ and at $z=1.5664$.}

\begin{figure}[h]
%\begin{center}
\centering
\includegraphics[height=7cm, width=7cm]{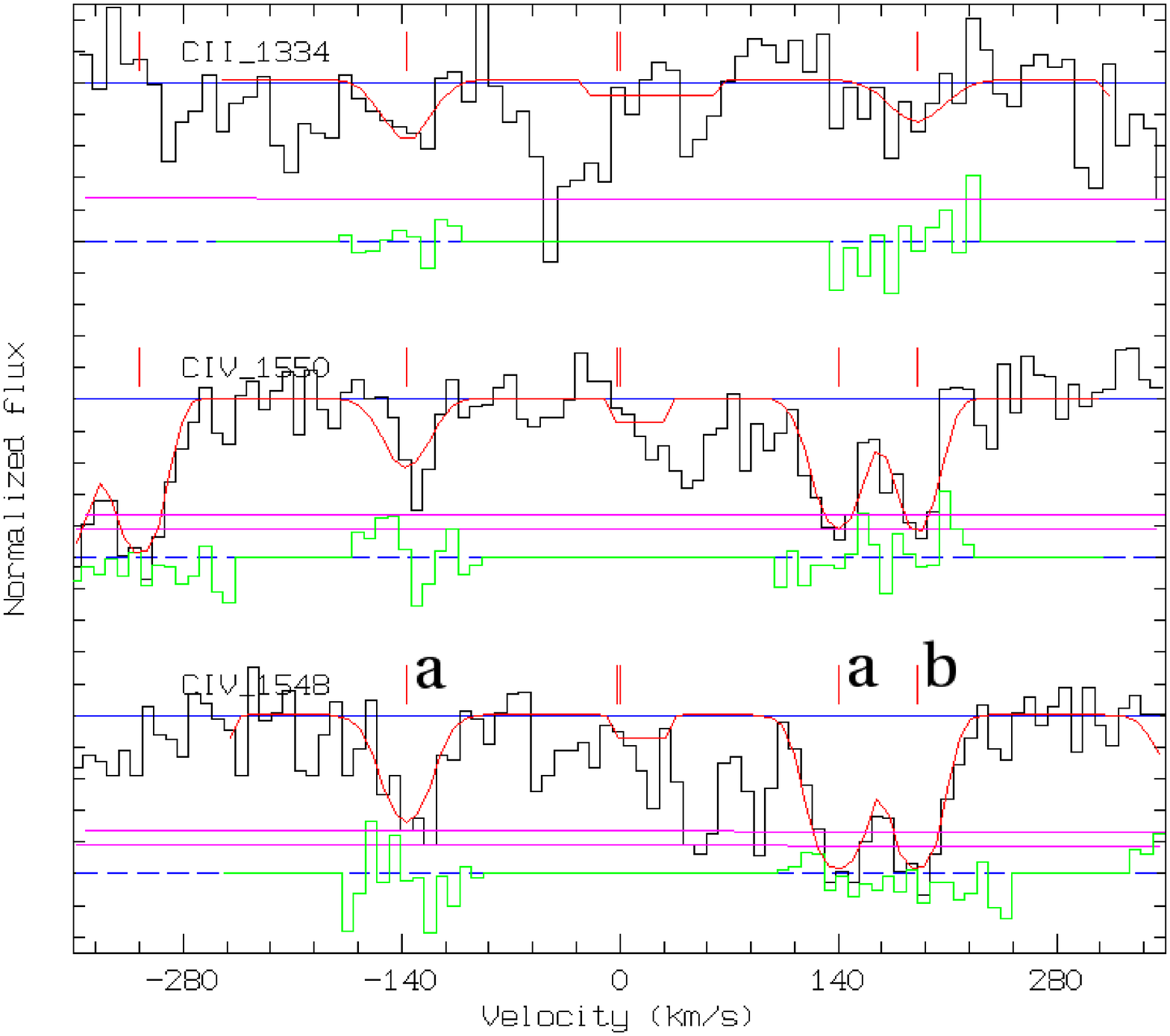}
%\end{center}
\caption{The intervening absorber at $z = 1.5664$ and $z =
1.5691$. The zero of the velocity scale refers to a redshift between
the two.}
\label{1_56}
\end{figure}

We discuss the intervening system at $z = 1.5691$ and $z=1.5664$
together, because the velocity shift between them is very small.  They show
high ionization line (CIV) and low ionization line (CII), (see Table
\ref{table_1_56} and Fig. \ref{1_56}).

\begin{table}[h]
\def~{\hphantom{0}}
{\tiny
\begin{center}
\caption{Column densities for the $z=1.5664$ and 1.5691 intervening systems.}
\begin{tabular}{ccc}
\hline
\bf Transition   &  1.5664a  \\
\hline
CII &13.6$\pm$0.2 \\
CIV &13.70$\pm$0.07 \\ 
\hline
\hline
\bf Transition   &  1.5691a & 1.5691b  \\
\hline
CII & - & 13.4$\pm$0.3 \\
CIV  &14.16$\pm$0.08& 14.1$\pm$0.1\\
\hline
\end{tabular}

\label{table_1_56}
\end{center}}
\end{table}

%%%%%%%%%%%%%%%%%%%%%%%%%%

\section{The $z=2.1992$ absorption system: physical properties of 
the GRB environment and host galaxy ISM}
\label{mainsystem}

\begin{figure}[h]
\centering
%\begin{center}
\includegraphics[height=10cm, width=10cm]{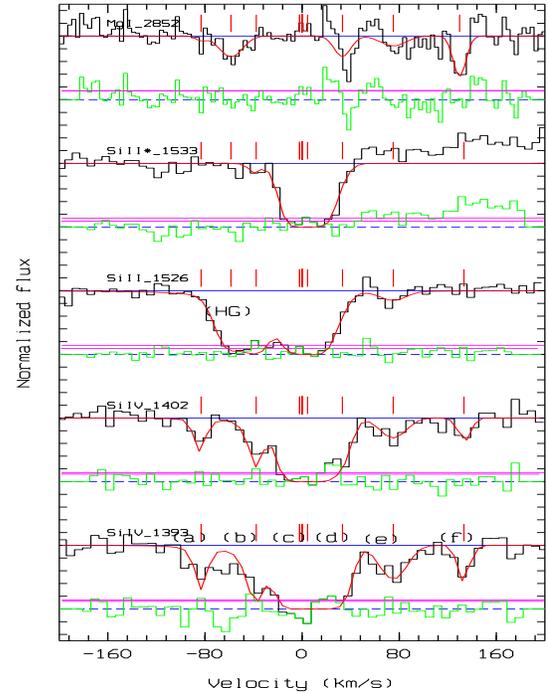}
%\end{center}
\caption{The UVES spectrum around the lines Si IV $\lambda$1393,1402,
SiII $\lambda$1526, SiII* $\lambda$1533 lines and
MgI$\lambda$2852. For MgI we clearly see the (HG) component.The zero
of the velocity scale refers to the redshift of the host galaxy,
$z=2.1992$.}
\label{abcdef}
\end{figure}

This system is most likely associated with the GRB host galaxy.
Likewise most GRB host galaxy absorption systems
\citep[e.g.][]{fiore05,delia07,vrees07,prochaska06, prochaska08, delia08} it presents many
components, spanning a rather large velocity range ($\sim210$
km\,s$^{-1}$). Interestingly, different components exhibit different
transitions: some exhibit low-ionization lines, high-ionization lines
and fine-structure lines, some exhibit low-ionization lines only.  An
example is given in Figure \ref{abcdef}. Six components, labeled from
(a) to (f), are identified for the SiIV and CIV doublets, spanning a
velocity range from -75 to +140 km\,s$^{-1}$. Each component has a
width from 10 to 25 km\,s$^{-1}$.  Component (c) has nearly zero
velocity shift and it may be associated with the GRB surrounding
medium, because of the strong \ion{Si}{II}*$\lambda1309$ the
high-ionization transitions and the fact that this component is not
detected in MgI.  Fainter, but still significant \ion{Si}{II}* lines
are associated with components (d) and (e) too. On the other hand, the
\ion{Si}{II}$\lambda1190$ component labeled (HG) as ``host galaxy'' at
$\sim -58.60$ km\,s$^{-1}$ is not present in either \ion{Si}{IV} and
\ion{Si}{II}* transitions, suggesting that this component is due to
some host galaxy cloud along the line of sight not modified by the GRB
radiation field.  For this component,  \ion{Mg}{I} is seen in
absorption (Fig. \ref{abcdef}), implying a distance from the GRB site
$\ga 100$~pc \citep{prochaska06}.

The complex line profile, and the fact that different components may
exhibit physically different line transitions, made it impossible to
fit the system with a single Voigt profile. Therefore we were forced
to identify the different components, before we could derive the
column densities through a fitting procedure.  The main goal of this
work is indeed to disentangle the relative contribution of the
different components.  We use the less saturated transitions to guide
this identification. We identified 7 components for the $z=2.1992$
system ((a) through (f), and (HG); see Fig. \ref{abcdef}).  The
redshifts of components (a) to (f), and therefore their velocity shifts
with respect to $z=2.1992$, which has been taken as a reference, were
determined using the CIV and SiIV lines.  The redshift of the (HG)
component was determined by using the SiII transitions. The redshifts
were taken fixed when fitting all other lines belonging to each
component with Voigt profiles. The Doppler parameters of these lines
were also linked together. The column densities of non saturated
and moderately saturated transitions of each component of the main
absorption system at $z=2.1992$ are given in Table \ref{column}.
Strongly saturated lines could not be used for column density
evaluation, and accordingly a  ``SAT" is given in the tables for the
corresponding transitions.

To test the robustness of the results, in terms of the accuracy and
stability, we performed many fits, using several
combinations of line components.  We found that the results presented
in Table \ref{column} provide a good compromise between increasing the
statistical precision of the fit, obtained by increasing the number of
components fitted simultaneously, and the stability/repeatability of
the results. The latter is degraded when the number of fitted
parameters is increased, because of the increasingly complex shape of
the $\chi^2$ hypersurface in the parameter space, which may contain
many local minima.  We verified that the total best-fit column density
of each system is stable, within the statistical errors, changing the
number of components in each system.

\begin{figure}[h]
\centering{
\includegraphics[height=10cm, width=10cm]{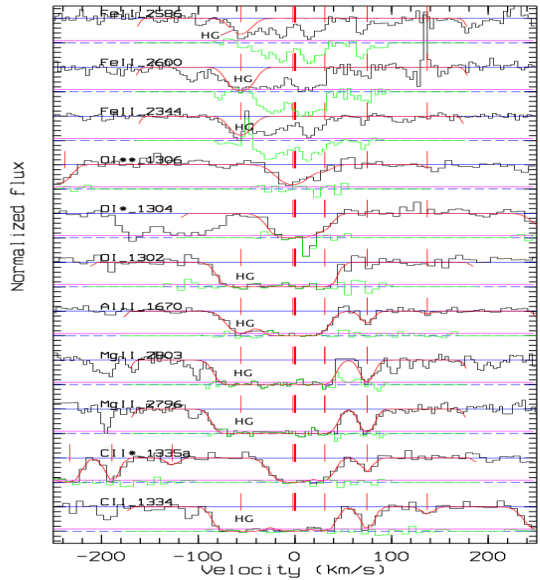}
}
\caption{The ``host galaxy'' component (HG) is present in all
low-ionization lines.  The column density determination for components
(a) and (b) is difficult due to their blending with the (HG) one.}
\label{hgcomp}
\end{figure}

%%%%%%%%%%

\begin{table*}
\def~{\hphantom{0}}
\begin{minipage}{150mm}
%\begin{center}
\caption{Derived column densities for the $z = 2.1992$ system.}
\label{column}
\begin{tabular}{cccccccc}
\hline
%%%%%%%%
\bf Transition   &  2.1992a & 2.1992(HG) & 2.1992b& 2.1992c & 2.1992d & 2.1992e & 2.1992f\\
\hline
km\,s$^{-1}$  &  -83.26 & -58.60    & -37.88& -0.39  &  +33.15& +77.54 & +136.73 \\ 
\hline
MgI  &  $\dag$             & 11.8$\pm$0.1&  $\dag$             & $<$ 11.8      & 12.0$\pm$0.1 & 11.6$\pm$0.2 &  $\star$ \\
OI   &  -             & 15.4$\pm$0.3&  -             & $\ddag$ & $\ddag$& -              & -             \\
OI*  &  -             &  -            &  -             & 15.1$\pm$0.1 & 14.0$\pm$0.2 &  $<$13.7             & -             \\
OI** &  -             &  -            &  -             & 14.53$\pm$0.08 & 13.4$\pm$0.3 & $<$13.7            & -             \\
%%%%%%
MgII &  $\dag$            & 14.6$\pm$0.3&  $\dag$            & 14.8$\pm$0.6 & 13.6$\pm$0.3 & 13.4$\pm$0.3 & 11.3$\pm$0.5   \\
CII  &  $\dag$             & 15.5$\pm$0.4&  $\dag$           & 14.6$\pm$0.2 & 14.6$\pm$0.5 & 14.2$\pm$0.3 & 13.06$\pm$0.21\\
CII* &  -             &  $<$ 13.3    &  -                  & 14.6$\pm$0.1 & 14.2$\pm$0.3 & 13.5$\pm$0.2 & $<13.3$            \\
SiII &  $\dag$             &14.26$\pm0.07$&  $\dag$             & $14.8\pm0.1$ & $13.9\pm0.2$ & $13.5\pm0.2$ & $12.9\pm0.2$\\
SiII*&  -             &  $<$ 13.0   &  -               & $14.28\pm0.04$ & $13.1\pm0.1$ & $13.1\pm0.1$ & $12.4\pm0.3$\\
AlII &  $\dag$             & 12.90$\pm$0.05&  $\dag$             & 14.0$\pm$0.3 & 12.1$\pm$0.2 & 12.15$\pm$0.07 & 10.6$\pm$0.5\\
AlIII&  $\dag$             &  11.8$\pm$0.1   & $\dag$ & 12.09$\pm$0.04 & 11.30$\pm$0.3  & -               & -               \\
%%%%11.94$\pm$0.2 AlIII (c)
%%%%%%
CIV  & 13.40$\pm$0.05 &   $<$ 13.26   & 13.15$\pm$0.09 & 15.0$\pm$0.2 & 13.63$\pm$0.07 & 13.78$\pm$0.03 & 13.12$\pm$0.07\\
SiIV & 13.07$\pm$0.07 &  $<$ 13.17   & 13.2$\pm$0.1 & 14.3$\pm$0.2 & 13.3$\pm$0.1 & 13.06$\pm$0.06 & 12.76$\pm$0.09\\
NV   &  -             &  -            & - & 14.06$\pm$0.08 & 15.3$\pm$0.7 & -              & -                      \\
%PV   &  -             &  -           &-  & 14.06$\pm$0.07 & $<$12 & -              & -                        \\
%
OVI  &  -             &  -            &- & 15.2$\pm$1.9 & 14.6$\pm$0.2 & -              & -                       \\
\hline
Fe   &  2.1992a & 2.1992(HG) & 2.1992b& 2.1992c(1) -- 2.1992c(2)&  2.1992d & 2.1992e & 2.1992f \\
\hline
km\,s$^{-1}$ & -83.26& -58.60        & -37.88         & -13.3 / +15.74                   &  +33.15         & +77.54          & +136.73 \\ 
\hline
FeII &  -    & 13.67$\pm$0.04&  13.75$\pm$0.03& 13.3$\pm$0.1 / 13.73$\pm$0.04 &     -           & 14.3$\pm$0.4  & -         \\
FeII*&  -    &  -            &  -             & -              /  13.36$\pm$0.07 &     -           & -  & -         \\
\hline
%%%%%%%%%%%
%%%%%%%%%%%%%%
\end{tabular}
\end{minipage}

%\end{center}
%}
\bigskip
$\dag$ This component is blended with the host galaxy component (HG), so
we cannot determine the relative column density.\\ $\ddag$ Saturated
component.\\ $\star$ Component confused by another one shifted to
5.5 km s$^{-1}$ with respect to (f) position.

\end{table*}

%%%%%%%%%%

\subsection{Low-ionization lines}

Strong low-ionization lines were detected for all components.  In four
cases (MgII, CII, AlII and AlIII) reliable estimates of the ion column
density for (c), (d), (e), (f) and (HG) components were possible.  In
all cases, strong blending of the (a) and (b) components with the (HG)
one renders such an estimate difficult for these components.  We
observe the SiII ($\lambda1190$\AA, $\lambda1193$\AA,
$\lambda1260$\AA, $\lambda1304$\AA, $\lambda1526$\AA), CII
($\lambda1334$\AA), MgII doublet ($\lambda2796$\AA, $\lambda2803$\AA),
AlII$\lambda1670$\AA{} and the AlIII doublet ($\lambda1854$\AA,
$\lambda1862$\AA).  From Table \ref{column} we can see that the host
galaxy component (HG) is present in all low ionization lines (see also
Fig. \ref{hgcomp}).

\subsection{High-ionization lines}

\begin{figure}[h]
\centering
%\begin{center}
\includegraphics[height=8cm, width=8cm]{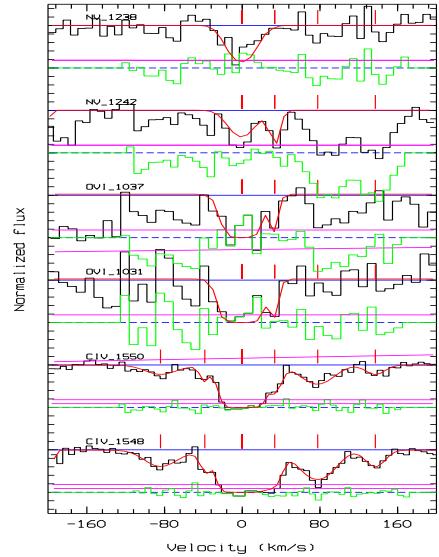}
%\end{center}
\caption{High-ionization absorption lines for the main system at
z=2.1992. The zero of the velocity scale refers to the redshift of the
host galaxy, $z=2.1992$.}
\label{highions}
\end{figure}

Figure \ref{highions} shows all high-ionization lines (NV doublet, OVI
doublet) identified together with SiIV and CIV. Significant column
densities of CIV and SiIV are detected for all six components (a) to
(f). NV and OVI are detected for components (c) and (d)
only. Confusion with the Lyman-$\alpha$ forest makes it difficult to
search for faint OVI and NV lines for the other components or to
compute robust upper limits.  No high-ionization lines were detected
for component (HG).

\subsubsection{The atoms MgI and OI}

We detect significant amounts of both neutral Mg and O.  The first
ionization energy of Mg lies below the 1~Ryd threshold, at 7.6
eV. This transition is not screened by H absorption and therefore Mg
would be ionized easily by the strong radiation coming from the
GRB. As a consequence, the presence of significant MgI implies large
distances ($>100$~pc, \citealt{prochaska06, Chen07}) of the gas clouds from
the GRB site.
MgI is detected for components (HG), (d) and (e) (see
Table \ref{column} and Fig. \ref{abcdef}). These components are therefore
likely to be associated with gas clouds which are little affected by
the GRB radiation field.  MgI of components (a) and (b) is
difficult to constrain because of the blending with the strong HG
component. MgI is not detected for component (c), which has very
high-ionization and fine-structure lines, and for component (f).  It
should be noted that a line is present at 5.5 km s$^{-1}$ from the position
of MgI at the redshift of component (f), a shift 
big for this line to be associate to the component.

%\begin{figure}[h]
%\centering

%\includegraphics[height=8cm, width=8cm, angle=-90]{silici_mg_NEW.eps}

%\caption{The UVES spectrum around the lines SiIV $\lambda$1393,1402,
%SiII $\lambda$1190, SiII* $\lambda$1194 lines and MgI
%$\lambda$2852. For MgI we clearly see the (HG) component.  The zero of
%the velocity scale refers to the redshift of the host galaxy,
%$z=2.1992$. }
%\label{MgI}
%\end{figure}

\subsubsection{Fine-structure lines}
\label{finestruct}

We detected several fine-structure lines associated with components (c),
(d), (e) and (f).  We identified the following lines arising from fine
structure levels: SiII*$\lambda$1194\AA, SiII*$\lambda$1197\AA,
SiII*$\lambda$1264\AA, SiII*$\lambda$1309\AA, 
SiII*$\lambda$1533\AA, CII*$\lambda$1335a\AA, OI*$\lambda$1304\AA, 
OI**$\lambda$1306\AA. The redshifts of all these lines were fixed
to those of the corresponding components in the CIV and SiIV systems.

\begin{figure}[h]
\centering
%\begin{center}
\includegraphics[height=8cm, width=8cm]{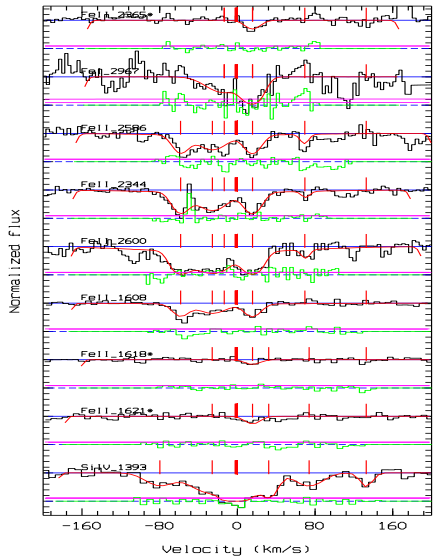}
%\end{center}
\caption{The UVES spectrum around some FeII and FeII* lines compared
with SiIV $\lambda$1393.  The zero of the velocity scale refers to the
redshift of the host galaxy, $z=2.1992$.}
\label{FeII}
\end{figure}

We also detect three FeII fine-structure lines: 
FeII*$\lambda$1618\AA, FeII*$\lambda$1621\AA,
FeII*$\lambda$2365\AA. The redshift of these lines is shifted from
that found for component (c) using the CIV and SiV systems by about 20
km\,s$^{-1}$, see Fig. \ref{FeII}. This suggests that additional
structure is actually present in this absorption system. In the
following we consider the FeII fine-structure lines as associated with
component (c) without making further splitting in velocity.  No FeII
fine-structure transition is found for the other six components and,
in particular, no fine-structure line was found at a redshift
consistent with that of component (HG).

\section{Properties of the medium surrounding the GRB} 

\subsection{Fine-structure line excitation I: constraining the density of the gas} 
\label{collisional}

As shown in section \ref{finestruct}, for this burst we found several
fine-structure transitions.  These levels can be populated (1)
indirectly through excitation by ultraviolet photons, followed by
fluorescence, (2) through collisions between the ion and other
particles like free electrons, (3) and/or via direct photo-excitation
by infrared photons.  The first mechanism has been conclusively shown
to be the preferred mechanism in two cases: GRB060418
\citep{vrees07} and GRB080319B \citep{delia08}. The latter mechanism is the less dominant one and we
will not consider it.  Whatever the case, the physical conditions of
the interstellar medium can be probed through the detection of
transitions from these energetically lower excited levels.

For GRB050922C fine-structure transitions are present in component
(c), (d), (e) and (f), (see Table \ref{column}).

If we assume that collisions with electrons are the dominant
excitation mechanism, we can put constraints on the electron density
of the components where we found such transitions.  A method to
estimate the electron density is presented by \citet{prochaska06}. We
can compare our results on the SiII*/SiII and FeII*/FeII ratios with
Fig. 9 of \citet{prochaska06}, showing collisional excitation of the
first excited state relative at the ground state for Si$^+$, Fe$^+$
and O$^0$.  We can only use {\ion{Si}{II}} and {\ion{Fe}{II}} because
oxygen is satured.  From this comparison we find for SiII $n_e$ in the
range $(1.3-3) \times 10^2$ cm$^{-3}$ for component (c), $n_e$ in the
range $(1.2-2) \times 10^2$ cm$^{-3}$ for component (d), $n_e < 2 \times
10^1$ cm$^{-3}$ for component (e) and $n_e > 2 \times 10^3$
cm$^{-3}$ for component (f).

We have evaluated the electron density of component (c) also using the
{\ion{Fe}{II}} transitions. This produces a result $n_e \sim (0.1-5)
\times 10^5$ cm$^{-3}$ inconsistent with the previous estimate.  A
possible reason for this inconsistency may be our simple description
of a complex system (the excited transition lines are shifted by 20
km\,s$^{-1}$ from the ground level lines).

\subsection{Fine-structure line excitation II: constraining the radiation field}
\label{UV_pumping}

A competitive mechanism responsible for the presence of fine-structure
levels is indirect UV pumping.  If this is the dominant excitation
mechanism, we can constrain the GRB radiation field and the distance
of the gas from the GRB. \citet{prochaska06} using the PopRatio code
developed by \citet{Silva01, Silva02}, analyzed the relation between
the far UV radiation field intensity and the relative fraction of
excited fine-structure states with respect to their ground levels (for
{\ion{O}{I}}, {\ion{Si}{II}} and {\ion{Fe}{II}}, see their Figs. 7 and
8). In the following we make use of these figures to estimate the
radiation-field intensity from our column density data.

Using Fig. 7 of \citet{prochaska06}, we find that our measured
ratio log(${\ion{Si}{II}}^*/{\ion{Si}{II}}) = -0.5\pm 0.1$ for component
(c) implies a radiation field intensity $G/G_0 = 2_{-1}^{+3} \times
10^5$ (90\% confidence interval; $G_0 =1.6\times 10^{-3}$ erg
cm$^{-2}$ s$^{-1}$ is the Habing constant).  We note that the
\ion{Si}{II} lines of this component are moderately saturated and
therefore that additional systematic uncertainty may affect their
column densities determination. Fortunately, for this component we
detect also \ion{Fe}{II} and \ion{Fe}{II}$^*$ transitions.  The
log(${\ion{Fe}{II}^*}/{\ion{Fe}{II}}$) is $-0.5\pm0.1$, implying a
radiation field intensity $G/G_0 = 6_{-5}^{+4} \times 10^6$,
consistent with the \ion{Si}{II} determination, within the rather large
statistical errors.

A similarly high value of $G/G_0$ is found for component (f) only,
for which we found log(${\ion{Si}{II}}^*/{\ion{Si}{II}}) = -0.5 \pm 0.3$,
implying a radiation field intensity of $G/G_0 = 3_{-0.4}^{+5} \times
10^5$.

For component (d) and (e) we found log(${\ion{Si}{II}^*}/{\ion{Si}{II}}) =
-0.8 \pm 0.2$ and $-1.4 \pm 0.4$ respectively, implying radiation
field intensity $G/G_0 = (8\pm1) \times 10^4$ and $G/G_0 =
1_{-0.7}^{+0.7} \times 10^4$, both significantly lower than for component
(c). We note that the \ion{Si}{II} lines of component (e) are certainly 
not saturated or blended with other components, while that of component 
(d) may be slightly blended and/or saturated.

While UV pumping is the likely mechanism for the population of fine
structure levels in this GRB spectrum, we cannot rule out collisional
exitation. The ``smoking gun'' of the UV pumping mechanism is time variability 
of fine structure lines \citep{vrees07,delia08}. Unfortunately we have only one single observation of
the GRB050922C and therefore we cannot look for line variations.
We note that components (c) and (f) experience the
highest radiation field.

\section{Discussion}

\subsection{Distances from the burst region}

\begin{figure}
\includegraphics[height=.45\textheight, angle=0]{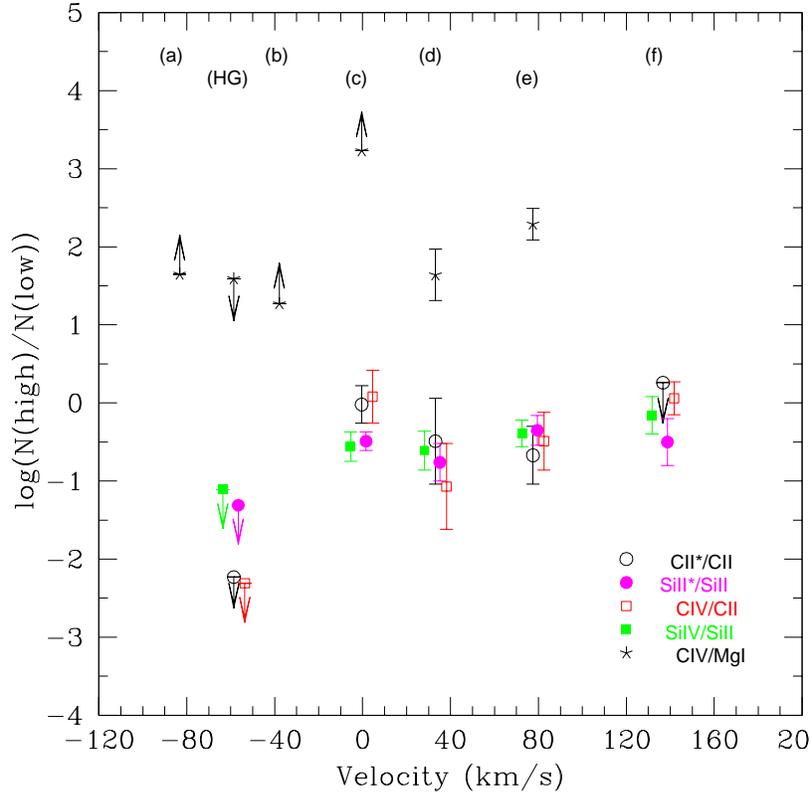}
\caption{Several high- to low-ionization ion column densities ratios for
the seven components of the $z=2.1992$ system as a function of the velocity shift
with respect to the redshift of the host galaxy. See text for details.}
\label{ionratio}
\end{figure}

We plot in Fig. \ref{ionratio} high to low-ionization line ratios and
fine structure to ground state line ratios for the seven components of
the $z=2.1992$ system. The different behaviour of component (HG) with
respect to all the others and in particular to components (c) and
(f) is evident from this plot. The most natural explanation for this
behaviour is to assume that the various components do not belong to
the same physical region in the host galaxy. The strength of the high
ionization lines and of the fine-structure lines, the absence of MgI
absorption, the analysis of the ratio of the excited levels to ground
levels and the metallicity analysis suggest that components (c) and (f) are closer to the GRB site than the other components.  
To better quantify this suggestion we performed detailed time evolving
photoionization calculations using the GRB light-curve to estimate the
photoionization fractions (assuming an optically thin medium). We used
the code initially developed by \citet{Nicastro99} and updated by
\citet{Krongold08}. We assumed a constant density profile
throughout the cloud, and a plane parallel geometry.  We studied gas
densities between 10 cm$^{-3}$ and $10^{8}$ cm$^{-3}$.  The ionizing
continuum was assumed to be a power law, $F(E)=E^{-\Gamma}$ photons
cm$^{-2}$ s$^{-1}$, with cutoffs at low and high energy. The high
energy cutoff was fixed at $10^{21}$Hz.

%OLD:Our calculations require that component (c) is closer than $\sim200$ pc to the GRB site, to reproduce
%the observed, quite large abundances of SiIV and CIV and the presence
%of OVI and NV. This is consistent with the results by \citet{prochaska08} 
%who suggest that the presence of NV implies a distance of $\sim10$ pc from the GRB site. This is also consistent 
%with \citet{Lazzati06} who put an upper limit of $\sim 150$ pc for the
%existence of these ions (though for models with metallicity 1000
%solar).

Our calculations require that component (c), in order to reproduce simultaneously
the lines by OVI, NV, CIV,  SiIV, FeII and OI, the gas density would have to be between 10$^{5}$ and 10$^{7}$ cm$^{-3}$, and the
distance between 200 and 300 pc from the GRB site\footnote{this density is implied only if both the high and low ionization lines come from the same gas phase.}

The presence of neutral elements in the spectra of GRB afterglows
can place strong constraints on the distance of the gas from the GRB.
For example, Prochaska et al. (2006) suggest that components showing
MgI absorption should be located at distances greater than 100~pc from
the GRB site. In the afterglow spectra of GRB050922C we detect neutral
MgI for three components: (HG), (d) and (e), see Fig. \ref{abcdef}.
In particular, the presence of strong MgI absorption, absence of
high-ionization lines and excited transitions, and low metallicity,
suggest that the (HG) component is located further away from the GRB
site, in a region of the host galaxy not strongly affected by the GRB
radiation field.  Our time evolving photoionization calculations
require a distance of at least 700 pc from the GRB site for this
component.  Components (d) and (e) do show fine-structure
transitions.  If the dominant excitation mechanism is UV pumping from
the GRB radiation field, this suggests a distance $\la 1$~kpc from the
GRB site \citep{prochaska07}. Our time evolving photoionization
calculations require a distances $\gs300$~pc for both components, to
be consistent with the observed C and Si line ratios.  Component (d)
exhibits also high-ionization absorption (NV, PV, OVI). The presence of these highly ionized
ions, along with the presence of low ionized ones (O I, Mg I, Fe II), can be explained at a distance
 of $\sim$ 300 pc from the GRB site if the density of this cloud is high enough, i.e. $>10^{5}$ cm$^{-3}$.

For components (a) and (b) we do not have information on the abundance
of neutral elements, because these transitions occur at wavelenghts
affected by the strong absorption of component (HG). The detection of
CIV implies a distance to the GRB site closer than $\sim700$ pc for both
components (a) and (b).

%\begin{figure}
%\includegraphics[height=.45\textheight, angle=0]{plot_UV_14_3.eps}
%\caption{The logarithm of the best fit ionization parameters for
%different components of [SiIV/SiII] ratios (filled squares) and
%[CIV/CII] ratios (open circles) as a function of the velocity shift
%with respect to the redshift of the host galaxy (upper panel: n=10
%cm$^{-3}$ and lower panel: $n=10^8 cm^{-3}$). Error-bars and upper
%limits represent the purely statistical $90\%$ confidence intervals.}
%\label{plotuv}
%\end{figure}

\subsection{Carbon and iron abundances}

Following \citet{delia07}, we calculated the average [C/Fe] ratio, an
estimator of the enrichment of the $\alpha$ elements relative to
iron. We measure a mean [C/Fe]$\mbox{} = 0.38 \pm 0.75$, consistent
with the value predicted by the models of \citet{pipino2004_I,
pipino2006_II} for a galaxy younger than 1 Gyr subject to a burst of
star-formation.  In such a case also a low [Fe/H] value is predicted,
close to that in Table \ref{tot_met_table}, and similar to what found by
\citet{delia07} for GRB050730.  Again, we note that mean abundance
estimates may be affected by large systematic uncertainties.

\subsection{Metallicity of the main system}

\begin{figure}[h]
\centering
%\begin{center}
\includegraphics[height=6cm, width=5cm, angle=-90]{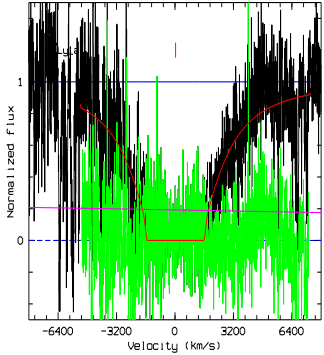}
%\end{center}
\caption{The Ly-$\alpha$ absorption feature.}
\label{lya}
\end{figure}

Figure \ref{lya} shows the Ly-$\alpha$ absorption features. Hydrogen
absorption results in very broad lines with damped wings, reaching
well beyond the velocity range of the heavier ions. Therefore is
clearly impossible to separate the seven components identified from
the analysis of \ion{C}{IV}, \ion{Si}{IV} and \ion{Si}{II}, and we can
only estimate the metallicity of the host galaxy in terms of the ratio
[X/H] between the total absorption column of the X element with
respect to that of hydrogen.

For this burst we measure a Hydrogen column density $\log N_{\rm HI} =
21.3 \pm 0.4$ cm$^{-2}$, similar to the average value found by
\citet{Jakobsson06} for a sample of Swift GRBs.

Table \ref{tot_met_table} gives, for each element, the total column
density of the seven components obtained by adding the column
densities for each ion (including ground states and the excited
fine-structure levels, if these are present) and the metallicity of the
element on a log scale. Taken at face value these metallicities are rather
small. Furthermore, a fraction of Hydrogen may well be ionized, in
particular that associated to the components closer to the GRB site
\citep{Waxman00, prochaska07}, suggesting even lower metallicities. On
the other side, the quoted metallicities are not corrected for dust
depletion. This is likely to be particuarly important for iron 
 \citep[see][]{Savaglio06} while it should be negligible for Silicon. The Si
metallicity in Table \ref{tot_met_table} is formally higher than that
of Fe, but statistically consistent within the large error bars. In
conclusion, even considering a little contribution from ionized
Hydrogen, we can exclude metallicities higher than $\approx 1/10$
solar.
 
This GRB was observed by the Swift X-ray telescope (XRT) Swift at the
same time of the UVES observations. It is therefore possible to
compare directly the column density determined from the Ly-$\alpha$
observation to that obtained from X-ray spectroscopy.  The spectrum
was fitted using model including a power law reduced at low energy by
intervening absorption at the redshift of the GRB host galaxy.  An
additional local absorber was included in the fit to account for
Galactic gas along the GRB line of sight. The column density of this
absorber was fixed at $5.75\times10^{20}$ cm$^{-2}$ \citep{Dickey90}.
Solar abundances were adopted for the Galactic absorber while the
metallicity of the GRB rest frame column density was fixed either to
solar metallicities or to 1/10 solar.  The best fit rest frame N$_H$
values are ($4.7 \pm 1.9)\times 10^{21}\mathrm{cm}^{-2}$, and ($3.0 \pm
1.0)\times 10^{22}\mathrm{cm}^{-2}$ for the two cases respectively. The latter value is higher than what found using the Lyman-$\alpha$ in the UVES spectrum a result consistent 
with what found by \citet{Jakobsson06} from other GRBs.

\begin{table}
\caption{Metallicity}
\tiny{
\begin{tabular}{lcc}
\hline
%\tableline\tableline
\bf Transition  &  $\log N^{(a)}$ (cm$^{-2}$) & [X/H] \\
\hline
{\bf HI} & $21.3 \pm 0.4$                  &  -                        \\
{\bf C}    &        $ 15.8\pm 0.5$ &    $ -1.9\pm0.5 $    \\
{\bf Si}    &         $ 15.3\pm 0.5$    &   $ -1.5\pm0.5 $    \\
{\bf O}      &     $ 15.7\pm 0.3$ &    $ -2.3\pm0.3 $    \\
{\bf N}      & $14.9\pm0.7$&             $   -2.3\pm 0.7$             \\
{\bf Al}       &  $14.2\pm0.5 $     &  $ -1.5\pm0.5 $    \\
{\bf Mg}     & $ 14.9\pm0.3 $     &  $ -1.9\pm 0.3$    \\
{\bf Fe}      & $ 14.9\pm0.4 $     &  $ -2.1\pm0.4 $    \\
\hline
\end{tabular}
}
\label{tot_met_table}
\normalsize

$^{(a)}$ Total column density of the element.
\end{table}

The metallicity estimates can be compared to metallicity derivations
that use optical emission lines from the GRB host galaxy. \citet{Savaglio06}
 found an average value of -0.15 for a sample of 11 GRB hosts at
z$<1$, a value significantly higher than those in Table \ref{tot_met_table}. 
The values in table \ref{tot_met_table} are significantly lower
than this feature. As discussed above several effects can play a role,
the main three being the exact correction for dust depletion, the
amount of ionized Hydrogen and the fact that we cannot estimate
the H column density of each single component. This is particularly
relevant, since different components can be associated with regions of
the host galaxy with very different metallicities. On the other hand,
emission line spectroscopy of GRB host galaxies is very difficult to
perform at $z>1$, meaning that at high redshift absorption line
spectroscopy of GRBs is probably the only tool to estimate host
metallicities.

\subsection{Metallicity of the intervening systems}

The Ly-$\alpha$ of three of the five intervening systems lies in the
UVES observed band. In these cases we were able to estimate the HI
column density (see Sections 3.2, 3.4 and 3.5), and therefore the
metallicity following the same approach used for the main system.

For the z=2.1416 intervening system we measure [C/H] =
$-1.2\pm0.1$. For the z=2.077 intervening system we measure [C/H]= -1.4, [Si/H]= -1.4, [Fe/H]= -1.3, [N/H]= -1.5, [O/H]= -1.4.
For this system we can also estimate the SiIV/CIV ratio, which is $-0.7\pm0.1$ for component (a) and $-0.8\pm0.2$ 
for component (b). These values are near the median value performed by \citet{songaila98}.
For the z=2.0081 intervening system we measure [C/H] = $0.7\pm0.5$, [Si/H] = $-0.1 \pm 0.4$ and
[Mg/H] = $-1.0\pm0.2$. The SiIV/CIV ratio, which is $-1.30\pm0.15$ for component (c), 
is similar to the median value computed by \citet{songaila98} but at
the lower end of the range found by \citet{dodorico01} for
systems at z$\sim1.9$. Finally, for the z=1.9885 intervening system we
measure [C/H] = $-0.1 \pm 0.5$.  These values are not corrected
for the fractions of Si and C in ions different from those in observed
transitions and reported in Tables 2, 4 and 5, and, most important are
not corrected for the fraction of H which may be ionized. This so
called `ionization correction' is a rather steep function of the
observed HI column density \citep{peroux07}, and for low observed
column densities, such those for the three intervening systems under
study, may be as large as -0.3 - -0.4.  Accounting for this ionization
correction, the [C/H] of the z=2.0081 and z=1.9885 systems
appear significantly higher than that found for the GRB host galaxy in
Table \ref{tot_met_table}. They are also higher than the typical
metallicity found in DLA and sub-DLA at similar redshifts \citep{peroux07}. We remark however that our determinations are based on rather sparse
line detections and therefore that a more systematic work is needed to investigate
further this issue.

\section{Conclusions}

Using the VLT high resolution spectrometer UVES, we have obtained
a high signal to noise ($\sim 20$) spectrum of the optical afterglow
of GRB050922C. Five intervening systems between $z = 2.077$ and $z =
1.5664$ have been identified along the GRB line of sight, in addition
to the main system at $z=2.1992$, which we identify as the redshift of
the GRB host galaxy.  The spectrum shows that the ISM of this galaxy
is complex, with at least seven components contributing to the main
absorption system. These systems show both high- and low-ionization
lines, fine-structure lines and neutral element absorption lines, thus
suggesting wide ranges of distances to the GRB site and physical
properties of the absorbing clouds. The analysis of ratios between
high and low ionization lines, the presence of strong fine structure
line and the analysis of the metallicity of the clouds allow us to put
quantitative constraints on the distance of the absorbing clouds to
the GRB site. In particular, we find that component (c) is likely to
be the closest component, with a distance $200-300$ pc from the GRB site.
Component HG is likely to be farthest component, with a distance $>700$ pc.  

We calculated the average metallicity of the GRB host galaxy by adding
the column densities of all seven components. This turned out to be between
1/100 and 1/10 solar, even for an element like Silicon, which should
not be strongly affected by dust depletion. This value is roughly
consistent with what is found by absorption spectroscopy in other
z=2-4 GRBs  \citep{Savaglio06, prochaska07} but is lower than the average found by
emission line spectroscopy at z$<1$ \citep{Savaglio06, Savaglio08}. Interestingly,
the C fraction of three intervening systems at z$\sim2$ is nearly
solar, and therefore significantly higher than that of the GRB host
galaxy.

The element transitions detected, the complexity of the components and
the column densities measured are similar to the other high redshift
GRB high resolution spectra present in the literature. 
We have no evidence of both the high velocity components reported for GRB021004 \citep{Mirabal03,
fiore05, Starling05}, GRB030226 \citep{Klose04} and GRB050505
\citep{Berger06}, and the strong asymmetry reported for GRB030329
\citep{Thoene07}.

\begin{acknowledgements}
We thank an anonymous referee for comments that improved the
presentation. We thank C. Porciani, V. D'Odorico, H.-W. Chen,
J. X. Prochaska and P.M. Vreeswijk for useful discussions.  Part of this
work was supported by MIUR COFIN-03-02-23 and INAF/PRIN 270/2003 and
ASI contracts ASI/I/R/039/04 and ASI/I/R/023/05/0.  SP and VD
acknowledge support from ASI grants.  SP, SDV and MDV thank the Dark
Cosmology Centre of Copenhagen, where part of this work was done, for
the friendly and creative atmosphere.  D.M. acknowledges the
Instrument Center for Danish Astrophysics for support and thanks the Dark Cosmology Centre (funded by the DNRF). SDV is
supported by SFI.
\end{acknowledgements}

\bibliography{paper_references.bib}

\end{document}